\documentstyle[11pt]{article}
\def\vsn{\vspace{2truemm}\noindent}

\begin{document}
\baselineskip=12pt

\begin{center}

{\bf Orientation of Galaxies in the Local Supercluster: A Review}


\vspace{0.5truecm}

F. X. Hu and G.X. Wu

\vspace{0.3truecm}
Purple Mountain Observatory, The Chinese Academy of Sciences, Nanjing 
210008,
and National Astronomical Observatories, The Chinese Academy of Sciences,
Beijing 100012

\vspace{0.3truecm}
Email:  fxyhhu@jlonline.com

\vspace{0.3truecm}
G. X. Song

\vspace{0.3truecm}
Shanghai Observatory, The Chinese Academy of Sciences, Shanghai 200030, and
National Astronomical Observatories, The Chinese Academy of Sciences,
Beijing 100012

\vspace{0.3truecm}
Q. R. Yuan

\vspace{0.3truecm}
Department of Physics, Nanjing Normal University, Nanjing 210024

\vspace{0.3truecm}
and

\vspace{0.3truecm}
S. Okamura

Department of Astronomy and Research Center for the Early Universe,
School of Science, University of Tokyo, Tokyo, 113-0033 Japan

\vspace{1truecm}
(Received\hspace{2truecm}; accepted\hspace{2truecm})

\end{center}


\centerline{\bf Abstract}
\vspace{0.3truecm}

The progress of the studies on the orientation of galaxies in the
Local Supercluster (LSC) is reviewed and a summary of recent
results is given.
Following a brief introduction of the LSC,
we describe the results of early studies based on two-dimensional
analysis, which were mostly not conclusive. We describe next
the three-dimensional analysis, which is used widely today.
Difficulties and systematic effects are explained and the
importance of selection effects is described.
Then, results based on the new method and modern databases
are given, which are summarized as follows. When the LSC is seen as a whole, 
galaxy planes tend to align perpendicular to the LSC plane with lenticulars showing
the most pronounced tendency. Projections onto the LSC plane
of the spin vectors of Virgo cluster member 
galaxies, and to some 
extent, those of the total LSC galaxies, tend to point
to the Virgo cluster center. This tendency is more pronounced for 
lenticulars than for spirals.
It is suggested that 'field' galaxies, i.e., those which do 
not belong to groups with more than three members, may be 
better objects than other galaxies to probe the information 
at the early epoch of the LSC formation through the analysis 
of galaxy orientations. 
Field lenticulars show a pronounced anisotropic distribution
of spin vectors in the sense that they lay their spin vectors 
parallel to the LSC plane while field spirals show an isotropic
spin-vector
distribution. 

\noindent
Key words: Local Supercluster, orientation, spin vectors, anisotropy

\section{Introduction}

Superclusters are the largest structure in the Universe with
scales up to $\sim$100--150 Mpc.
They are in largely unrelaxed state. At a typical velocity of $\sim$1000
km s$^{-1}$, galaxies can move only $\sim$10 h$^{-1}$Mpc within the Hubble
time (Hubble constant $H_0$=100h km s$^{-1}$ Mpc$^{-1}$).
Accordingly, the superclusters seen today tell us the condition of the Universe
when these largest structures were formed (Oort, 1983; Bahcall, 1988).
The dynamical 'fossil' nature of the alignment of rotational axes of
galaxies in superclusters provides us with the information on the 
formation process of galaxies at the early epoch (Djorgovski, 1987).

The orientation of galaxies in the Local Supercluster (hereafter LSC)
has been studied extensively by many investigators
(e.g., Reinhardt and Roberts, 1972; Jaaniste and Saar,
1977, 1978; Kapranidis and Sullivan, 1983 (hereafter KS);
MacGillivray et al., 1982; MacGillivray and Dodd, 1985a,b; 
Flin and Godlowski, 1986 (hereafter FG);
Kashikawa and Okamura, 1992 (hereafter KO); Godlowski, 1993, 1994;
Garrido et al., 1993; Han et al., 1995; Sugai and Iye 1995;
Yuan et al., 1997; Hu et al., 1998; Aryal and Saurer, 
2001a, 2005c).
However, the results of these studies are quite uncertain, and
some of them are even in contradiction with each other.
Some authors found that the galaxy planes tend to be parallel to
the LSC plane, while others found no strong tendency
of alignment, or even showed that galaxy planes tend to be
perpendicular to the LSC plane.

In this paper we attempt to clarify the cause of the uncertainty and
summarize what we have learned on the orientation of galaxies in 
the LSC. We will put our primary emphasis on disk galaxies, i.e., spirals
and lenticulars, and the global features of galaxy alignment of 
the LSC as a whole.
This is because disk galaxies are rotationally supported, and 
their spin vectors are the dynamical parameter related to the
formation processes at early epoch, and the global features of galaxy
alignments of the LSC will shed some light on the origin and 
formation of the LSC.
Discussion of the results of many studies on the orientation of
galaxies in other groups, clusters, and superclusters
(e.g., Dojorgovski, 1987; Han et al. 1995; Cabanela and Aldering, 1998;
Fuller et al., 1999;  Flin, 2001; Aryal and Saurer, 2001b,
2004b, 2005a, b;
Bukhari and Cram 2003, etc.)
is beyond the scope of this paper.

We start in section 2 with a brief introduction of the LSC.
Early studies are reviewed in section 3.  The new three-dimensional
analysis and the statistical methods used are described in section 4.
Early results from the three-dimensional method are summarized in section 5.
Difficulties and systematic effects are discussed 
in section 6.
The results obtained with the new method applied to disk 
galaxies selected from modern databases are described in section 7.
Finally, a summary and prospects are given in section 8.

\section{The Local Supercluster}

The earliest recognition of the LSC dates back to the William
and John Hershel's work, who showed an excess of bright nebulae
in the northern galactic hemisphere (Herschel, 1784, 1785, 1802, 1811).
More recent evidence for its existence was given by Reynolds and 
Lundmark
in 1920's (e.g., Lundmark, 1927), Holmberg (1937) and G. de Vaucouleurs 
(1953, 1956, 1958, 1960, 1978, 1981).

De Vaucouleurs was the first who presented the firm observational
evidence for the LSC. He pointed out that the aggregation of the brightest
galaxies in the northern galactic hemisphere which extends along the
great circle is a flat super-system (Supergalaxy) with the axial ratio about
5:1 centered on the Virgo Cluster. He also defined some fundamental 
parameters of the LSC. The position of the supergalactic pole is defined
as $\alpha$(1950)=18$^h$52.$^m$8, $\delta$(1950)=+15$^{\circ}38'$.
Supergalactic coordinates (L and B) are introduced by de Vaucouleurs et al.
(1976) (see also MacGillivray et. al., 1982).
The center of the Local Supercluster is in the direction of the Virgo
Cluster, $\alpha$(1950)=12$^h$28$^m$, $\delta$(1950)=+12$^{\circ}40'$,
and its distance is D =15 Mpc (de Vaucouleurs, 1982).
The position of the Virgo Cluster is in the direction of L=103$^{\circ}$, and 
B=$-2^{\circ}$. A review paper of the LSC was presented at the IAU Symposium 
No.79 (de Vaucouleurs, 1978).

A detailed description of the global structure of the LSC based on galaxy 
space coordinates, derived from their radial velocities, can be found in
Yahil et al. (1980), Oort (1983), Tully (1982, 1988) and the references
therein. With the present knowledge, about one-third of the mass of the LSC
resides in the halo component and the remaining two-thirds are in the
disk component, with about a half of its mass contributed by the Virgo 
cluster. About 70\% of nearby galaxies belong to groups or clusters within the 
LSC if the Virgo members are included.

Some catalogs relevant to the LSC include:
\begin{itemize}

\item The Shapley and Ames Catalog (hereafter SA) \\
A catalog of 1249 galaxies brighter than 13.0 photographic magnitude
published by Shapley and Ames (1932) based on the Harvard photographic
survey. They revealed the existence of the ``Super-galactic Equator''
(see also Oort, 1983), which is the aggregation of bright galaxies along
the great circle that passes through the Virgo Cluster, the core of the LSC
as we know now.

\item The Revised Shapley and Ames Catalog of Bright Galaxies (hereafter RSA) 
\\
A revised catalog of SA published by Sandage and Tammann (1981, 1987),
which contains magnitudes, morphological types, and redshifts of 1246
galaxies in the SA. This is the first catalog that gives galaxy
{\it distances} based on the redshift and a Virgo infall model.

\item The Nearby Galaxies Catalogue (hereafter NBG) \\
A catalog of 2367 galaxies with velocities
smaller than 3000 km s$^{-1}$ as of 1978 (Tully 1988). Among the 2367 
galaxies,
1053 belong to RSA and 1515 come from all sky HI survey by the author and
collaborators. NBG is a companion to the Nearby Galaxies Atlas (Tully and
Fisher, 1987).

\item Photometric Atlas of Northern Bright Galaxies (hereafter PANBG).\\
A uniform detailed surface photometry database of 791 galaxies
in the northern hemisphere selected from the RSA published
by Kodaira et al. (1990). It contains photometry parameters, isophotal
contours, and luminosity profiles measured in the $V$ band.

\end{itemize}

\section{Early Studies Based on Two-Dimensional Analysis}

The study of galaxy orientation  in the LSC can be traced
back to long time ago (e.g., Reynolds, 1920, 1922, Holmberg, 1937;
Danver, 1942; Brown, 1938, 1939, 1964, 1968; Thompson, 1973;
see also Hawley and Peebles, 1975
and references therein). The advent of the Uppsala General Catalogue of
Galaxies (Nilson, 1973; hereafter UGC) and the ESO/Uppsala Survey
of the ESO(B) Atlas (Lauberts, 1982; hereafter ESO/U), which 
give the
position angles for thousands of galaxies, stimulated the research 
in this field.

Results of earliest studies were inconclusive.
Reynolds (1920, 1922) showed that if the shape of spiral galaxies
is approximated by a thin disk, there is an excess of large spirals
at small inclination angles to the line of sight.
Using 593 galaxies with diameters greater than $2'$ in the catalog of
Herschel nebulae, Brown (1938) found an apparent large excess of small
inclinations (estimated by axial ratios $b/a$) to the line of sight, which he
believed to be a real phenomenon due to a systematic orientation
of galaxy planes in space.

From the analyses of distribution of position angles of galaxies in 
Shapley's catalog of 7889 external galaxies in Horologium and 
surrounding regions and SA, Brown (1939, 1964, 1968) found evidence for
anisotropy in the orientation in some sky regions,
while Reaves (1958) found no such effect by measuring position angles
of galaxies on the two plates which cover in part the region of the
Shapley's survey. Kristian (1967) was also unable
to find anisotropy. Hawley and Peebles (1975)
measured orientations of 5559 galaxies in three sky areas on the
Palomar Observatory Sky Survey (hereafter POSS) red 
plates and found
no significant deviation from anisotropy.

The results of later studies did not improve the situation very much.
Reinhardt and Roberts (1972) studied the mean apparent ellipticity of
galaxies in the Reference Catalogue of Bright Galaxies (de Vaucouleurs and 
de Vaucouleurs, 1964, hereafter RC1) and concluded that orientations
of the spin vectors (hereafter SVs) of the galaxies are preferentially
perpendicular to the LSC plane. From a study of 1027 galaxies within 25 Mpc
(radial velocity $V<2500$ km s$^{-1}$ or $m<12.5$) in RC1 and UGC, 
Jaaniste and 
Saar (1977) found the opposite trend with the spin vectors parallel to the LSC 
plane.
In 1953-54, de Vaucouleurs studied the position angles of galaxies
close to the supergalactic equator in 
the catalog by Reinmuth (1926) and
examined the distribution of the poles of the 200 large spirals 
in the catalog by Danver (1942).
De Vaucouleurs (1981) claimed in an abstract that
this
unpublished 
research failed to detect any strong tendency
of the galaxy planes to be parallel, i.e. the SVs were perpendicular,
to the LSC plane.
MacGillivray et al. (1982) analyzed the distribution of
the position angles and ellipticities of 727 spirals and irregulars with
$B_{T}<13.0$ mag. and radial velocity corrected for solar motion $V_{0}$ $<$ 2500 km 
s$^{-1}$
in the Second Reference Catalogue of Bright Galaxies
(de Vaucouleurs, G. et al., 1976; hereafter RC2) and the UGC.
They found that galaxies belonging to the LSC have a slightly non-random 
distribution of their SVs. The SVs tend to be perpendicular to the LSC plane. 
This trend is most pronounced for galaxies at high supergalactic latitude and 
those seen nearly edge-on.

In summary, the results of these early studies were inconsistent, 
and at times even in contradiction with each other. 
Some authors concluded that the galaxy planes tend to be parallel to the LSC 
plane, i.e. their SVs are perpendicular to the LSC plane (e.g., Reinhardt and 
Roberdts 1972; MacGillivray et al. 1982). 
On the other hand, 
others concluded that galaxy
planes tend to be perpendicular to the LSC plane, i.e. their SVs are parallel
to the LSC plane (e.g., Jaaniste and Saar, 1977, 1978). 
Finally, some
authors found no strong tendency of alignment at all (e.g., KS). 
Possible causes for these inconsistencies include probably 
different data sampling, large errors 
in the data, selection effects, background contamination, and 
incompleteness of 
the sample, etc.

It might be worthwhile to mention a possible evolutionary effect on the galaxy 
orientation. Using the III$_a$J plates taken with the UK 1.2 m Schmidt
Telescope, MacGillivray and Dodd (1985b) examined
the position angle distribution of galaxies in a magnitude-limited sample 
down to $B$=16 mag in the Virgo region (25 square degrees). 
They did not find a global effect of alignment of Virgo galaxies 
along the LSC plane.
However, the galaxies did show  a strong alignment in the sense that
the galaxy planes are preferentially oriented
perpendicular to the radius vector to NGC 4406 near the Virgo center.
They concluded that this perpendicularity is unlikely to be a primordial
effect. The most likely cause 
is a dynamical effect
in the post-formation epoch, such as rotation around the Virgo center,
infall to the center, or tidal interaction between galaxies and the massive
cluster core. Using a new method and a new database (see below), KO found 
a similar effect. They concluded that the projections on the LSC plane of the SV 
of the galaxies in the core and the vicinity of the Virgo cluster tend to
point towards the Virgo center. This effect was also confirmed by
Hu et al. (1995; hereafter HWSL).

The method generally used in early studies was to look for anisotropy
in the distributions of position angles $PA$ and/or 
axial ratios $b/a$ of galaxy images. The position angle distribution and the 
axial ratio distribution
were examined {\it independently}. It was merely a two dimensional
analysis in nature which deals only with the projection of galaxies on the sky.
Since we are inside the LSC, the projected shapes of galaxies look different
depending on the direction and the distance from us. Therefore, if any
alignment of SVs in the three dimensional space exists it may be diluted
by this projection effect.

\section{Three-Dimensional Analysis and Statistical Tests}

A three-dimensional analysis, which analyzes position angles and
axial ratios of galaxies {\it simultaneously}, was introduced as a new method
in late 70's
(Jaaniste 
and Saar, 1977, 1978; KS; FG).

Before we describe this method frequently applied now, we will clarify first the 
coordinate system in which spatial orientation
and distribution of the galaxies of LSC can be specified.
The position of a galaxy is expressed in terms of the supergalactic
polar coordinates (L, B, distance) or the supergalactic Cartesian
coordinates (SGX, SGY, SGZ). The latter was introduced by Tully (1982).
The SGZ axis was defined to point to the direction B=+90 $^{\circ}$; the
SGX axis was aligned toward L=0 $^{\circ}$, B=0$^{\circ}$; and the
SGY axis was aligned toward L=90$^{\circ}$, B=0$^{\circ}$.
In the supergalactic Cartesian coordinate system defined above,
the SGX-SGZ plane is almost coincident with the plane of the Galaxy;
the positive SGY axis is only 6$^{\circ}$ off the north Galactic pole.
The direction of the Virgo cluster center is close to the SGY axis
(offset by 13$^{\circ}$).

To specify the orientation of the SV of
a galaxy, two angles are defined. The angle $\theta$ is the polar
angle between the SV and the LSC plane, and the angle $\phi$ is the
azimuthal angle between the projection on the
LSC plane of the SV and the SGX axis. A schematic illustration of
angles $\theta$ and $\phi$ is shown in 
Fig. 1, which is taken from KO.
In FG and some other papers (e.g. Hu et al., 1998), the basic great circle
`Meridian' of the supergalactic system is selected to go through
the center of the Virgo cluster: $\alpha$ = 186.$^{\circ}$25, and
$\delta$ = +13.$^{\circ}$10 (Sandage and Tammann, 1976).

The required observational data are the position angle $PA$
and the axial ratio $b/a$ of a galaxy. The inclination 
angle $i$ of the galaxy 
can be derived from the observed axial ratio $b/a$ and the intrinsic 
flatness, i.e., true axial 
ratio 
$q_0$, which is dependent on the 
morphological type. The values of $q_0$ for different types can be 
taken from  Heidmann, Heidmann and de Vaucouleurs (1971) or 
Bottineli, Gouguenheim, Paturel, and de Vaucouleurs (1983). 
It is sometimes assumed that $q_0$ is constant, say 0.2 
(Holmberg, 1958; Tully, 1988), for disk galaxies of all types.
Through the $PA$ and $i$, the spatial orientation of the 
SV of 
a galaxy 
is specified by angles $\theta$ and $\phi$.

In practice, the transformation of galaxy position between the equatorial
and supergalactic coordinate systems should be established first.
Based on the position angle measured with respect to the equatorial 
coordinate system $p$, the position angle with respect to the 
supergalactic coordinate system $PA$ can be obtained 
(MacGillivray et al., 1982; KO).
The angles $\theta$ and $\phi$ that specify the SV orientation of the
galaxy can be derived from the following formulae given by FG and KO:
\begin{equation}
{\rm sin}\;\theta=-{\rm cos}\;i\:{\rm sin}\;B\pm{\rm sin}\;i\:{\rm 
cos}\;q\:{\rm cos}\;B,
\end{equation}
\begin{equation}
{\rm sin}\;\phi=({\rm cos}\;\theta)^{-1}\left[-{\rm cos}\;i\:{\rm 
cos}\;B\:{\rm sin}\;L
+{\rm sin}\;i\:(\mp\;{\rm cos}\;q\:{\rm sin}\;B\:{\rm sin}\;L\pm{\rm 
sin}\;q\:{\rm cos}\;L)\right],
\end{equation}
\noindent
where $i$ is the inclination angle estimated from the axial ratio by
\begin{equation}
{\rm cos}^2 i=[(b/a)^2 - q_0^2]/(1-q_0^2),
\end{equation}
\noindent
which is valid for oblate spheroids (Holmberg, 1946).
The angle $q$ is defined as $q=PA-\pi/2$.

There is one limitation in this analysis. As KO pointed out,
there are four solutions for the orientation of 
the SV of a galaxy with a given position angle and a given axial ratio. 
This ambiguity is caused by the fact that we do not know which side 
of the galaxy is closer to us,
and whether the spin axis is pointing towards or away from us.
Usually all these four possibilities are counted as independent entries.

With the optical image and HI velocity fields, it is possible to
solve the ambiguity and therefore determine the orientation of the
SV of a galaxy uniquely.
By using the HI velocity field, we can decide which side of a
galaxy is approaching towards us. The optical image can tell 
us the direction of
the SV if we assume that spiral arms are trailing.
Helou and Salpeter (1982) and Hoffman et al. (1989) examined the
orientations of the SV of Virgo galaxies by this method.
Unfortunately, however, the available sample was too small to derive
a firm conclusion. Cabanela and Dickey (1999) used HI observations
with Arecibo telescope to determine the SVs of galaxies in the
Pisces-Perseus supercluster. Based on the analysis of the SVs of 54 nearly
edge-on galaxies, they found no significant evidence for alignment.
The sample size of this study is, however, also not 
large enough.

We give below a brief introduction on the statistical methods used
to evaluate whether the observed SV distribution is isotropic or not.
There are several kinds of statistical tests.
Here we discuss the chi-square ($\chi^2$) test and the Kolmogorov-Smirnov
test (hereafter K-S test), which 
are commonly used. Some other statistical 
tests, say, the Fourier test and
the autocorrelation test, were also used and discussed in detail
by FG and Godlowski (1993, 1994).

\vspace{3mm}\noindent
1) The $\chi^2$ test

The  $\chi^2$ test is an objective way to estimate the goodness
of the fit (e.g., to the isotropic distribution) based
on the reduced chi-square value, $\chi_{\nu}^2$, which
is defined by

\begin{equation}
\chi_{\nu}^2= \frac{1}{\nu}\chi^2=\frac{1}{\nu}\sum_{i=1}^n
\frac{(N_{oi}-N_{ei})^2}{N_{ei}},
\end{equation}
where $n$ is the number of bins,
$\nu=n-1$ is the degree of freedom, and $N_{oi}$ and $N_{ei}$
represent the observed and expected (isotropic) values in the
$i$th bin, respectively.  The quantity
$P(>\chi^2_{\nu})$ defined by
\begin{equation}
 P(>\chi^2_{\nu}) = 1-\big[2^{\nu/2}\Gamma(\nu/2)\big]^{-1}\int_0^{\chi^2}
e^{-\frac{t}2}t^{\frac{\nu}2-1}dt,
\end{equation}
 and gives the probability that the observed
distribution is realized from the isotropic distribution.
 The observed distribution is
more consistent with the isotropic distribution when
$P(>\chi^2_{\nu})$ is larger. For the isotropic distribution,
the $P(>\chi^2_{\nu})$ value is expected to be nearly 1.
The critical value $P(>\chi^2_{\nu})
=0.05(5\%)$ is often set to discriminate between isotropy and
anisotropy, which corresponds to the significance at
$2\sigma$ level. Usually  10$^\circ$  is adopted as the
bin size for the observed
$\theta$ and $\phi$ distributions.

There is a limitation for the $\chi^2$ test that data 
must be binned, and therefore some information may
be lost 
by binning.
The K-S test, which can be applied without any binning, does not
suffer from this disadvantage.

\vspace{3mm}\noindent
2) K-S test

Let the cumulative distribution function of an observed sample
$(x_1,x_2,...,x_n)$ be $S_n(x)$, and the known theoretical (e.g., isotropic)
one be $P(x)$. K-S test is based on the maximum absolute difference 
between these two cumulative distribution functions defined by
\begin{equation}
 d=\max \limits_{-\infty < x < +\infty}^{} \vert S_n(x)-P(x)\vert.
\end{equation}
The probability $\alpha$ that $S_n(x)$ deviates from $P(x)$ 
is approximated by
\begin{equation}
\alpha(>d)=Q_{KS} \bigl( [\sqrt{n}+0.12+0.11/\sqrt{n}]d \bigr),
\end{equation}
where
\begin{equation}
Q_{KS}(\lambda)=2\sum_{j=1}^{\infty} (-1)^{j-1}e^{-2j^2\lambda^2}.
\end{equation}
$Q_{KS}$ is a monotonic function with the limiting values: $Q_{KS}(0)=1$,
$Q_{KS}(\infty)=0$. The observed sample does not follow the theoretical
distribution function $P(x)$ when $\alpha(>d)$ is very small. Since it is
a test to examine whether or not two distributions are drawn from the same
population, the K-S test can be used to judge whether the SV distributions
of spirals and lenticulars follow the same distribution function or not as
well as to judge the isotropy.

The SV's polar angle $\theta$ ranges from $0^\circ$ to $90^\circ$,
and the azimuthal angle $\phi$ from $-90^\circ$ to $90^\circ$. For
the isotropic SV distribution, the expected cumulative
distribution functions are 
\begin{equation}
P(\theta) =
\int_0^{2\pi}\int_0^{\theta}r^2 cos\theta d\theta
d\phi/\int_0^{2\pi}\int_0^{\pi/2}r^2cos\theta d\theta d\phi = sin
\theta,
\end{equation}
and
\begin{equation}
P(\phi) = \frac12+\frac{\phi}{\pi},
\end{equation}
respectively.

\section{Early Results from the Three-Dimensional Analysis}

KS analyzed various samples of spiral galaxies of the LSC
from UGC ranging in size from 425 to 1154
and found no strong tendency of alignment.
However, their high-latitude sample No.2 (702 galaxies with Zwicky
magnitude $m_{z}\leq14.0$ and $|B|>10^\circ$ ) showed
a marginal sign that galaxy planes align parallel to the LSC plane
(2$\sigma$-effect).

On the other hand, FG analyzed the data of 1275 galaxies in UGC
which have radial velocities smaller than 2600 km s$^{-1}$ corrected for
solar motion. Following Hawley and Peebles (1975),
FG used three statistical methods, which are the $\chi^{2}$ test,
the Fourier test, and the autocorrelation test.
The value 45 $^\circ$ was defined as the boundary between the
direction parallel to and that perpendicular to the LSC plane.
FG found a tendency that galaxy planes align perpendicular to
the LSC plane, which is observed mostly for face-on galaxies whose
distribution is different from that of edge-on galaxies.
They also found a weak tendency that the projection of SVs
of galaxies onto the LSC plane tend to point to the Virgo center.
Due to the lack of data, KS and FG both assumed
0$^{\circ}$ for the position angle of face-on galaxies.
FG argued that (1) a significant fraction of non-LSC members
are present in KS's sample No.2 and that (2) the marginal tendency
seen in KS's sample No.1 (425 galaxies with galactocentric redshift
$V_0 < 2500$ km s$^{-1}$),
which KS did not regard as significant, is consistent with
their finding that galaxy planes align perpendicular to the LSC plane.

Flin and Godlowski (1989) and Godlowski (1993, 1994) analyzed a sample
of 2227 galaxies taken from UGC and ESO/U, in both northern
and southern hemispheres, which have radial velocities smaller
than 2600 km s$^{-1}$ corrected for solar motion and
another sample from NBG.
They found that the distribution of galaxy planes throughout the LSC is
anisotropic; planes tend to align perpendicular to the LSC
plane. They also found that SVs of galaxies projected on the LSC plane
show a tendency to point to the center of the Virgo cluster.
These effects were shown to be more pronounced for the 'non-spiral'
galaxies (all types other than spirals and lenticulars)
than for the 'spiral' galaxies (lenticulars and spirals).
Figure 2 is taken from Godlowski (1994), which shows the observed
(solid line) and the theoretical isotropic (dashed line)
distribution of the angle $\delta$(= $\theta$  
in this paper) (a),
and angle $\eta$ (=$ \phi$ 
in this paper) (b), 
for the whole
sample of galaxies.
Flin and Godlowski (1989) concluded that the parallelism of galactic planes
to the LSC plane found previously by some authors was due to the
exclusion from the analysis of face-on galaxies whose $PA$ is 
difficult to be determined.

Godlowski (1994) found that galaxy alignments strongly dependent
on supergalactic coordinates,
radial velocity distance and galaxy structure.
Planes of galaxies with small $|SGB|$ (at low supergalactic latitudes)
tend to be perpendicular to the LSC plane while planes of
galaxies with large $|SGB|$
(at high supergalactic latitude) tend to be parallel to the LSC plane.
He claimed that the detected perpendicularity of galactic planes
to the LSC plane might be caused by the perpendicularity of the
galactic planes to the radius vector pointing to the LSC center.

Parnovsky et al. (1994) analyzed edge-on galaxies in UGC, ESO/U, and
their own catalog (Karachentsev et al., 1993) and detected a statistically
significant anisotropy of galaxy orientations with an excess of
projected SVs pointing to the region $4^h<\alpha<6^h$ and
$+20^{\circ}<\delta<+40^{\circ}$. Not all the sample galaxies had
redshift information and they claimed that the observed anisotropy
is global over the scale of 100-200 Mpc, based on the analysis of
apparent size of galaxies. However, Flin (1995) indicated that
this anisotropy is due to the LSC.

It is mentioned here that the three-dimensional method, when
face-on galaxies are included in the analysis, began to give
a consistent result that the planes of LSC galaxies tend to
align perpendicular to the LSC plane and  that SVs of galaxies
projected on the LSC plane tend to point to the
center of the Virgo cluster. (e.g., FG, Goldlowski, 1993, 1994,
Flin 1996).  However, there are two issues of concern
which should be mentioned. First, majority of the studies
mentioned above used 'all galaxies' as sample galaxies, i.e.,
ellipticals, lenticulars, spirals, irregulars and others.
Second, most of the studies after 1973 used the UGC data, which
are now known to suffer from selection effects.
In the next section we give a discussion about technical problems
related to these issues, i.e., difficulties and systematic
effects encountered in determining galaxy orientations.

\section{Difficulties and Systematic Effects in Studying Galaxy Alignment}

\subsection {Position Angle and Axial Ratio}

\noindent
(a) Selection Effect of UGC

\vspace{2mm}
There are many factors that can affect the determination of 
position angle $PA$ and axial ratio $b/a$. We discuss first the selection 
effects in the UGC, since most studies after 1973 used this catalogue.  

The UGC catalog was compiled by visually inspecting the galaxies 
larger than 1 arcmin on the POSS blue 
and red prints. The position angles and axes of galaxies were 
measured by eye. As it is more difficult to measure $PA$ for 
rounder galaxies, this leads to larger errors of $PA$
for rounder galaxies. 
For the face-on galaxies 
the position angles are not given in the UGC.

KO compared $PA$ and $b/a$ of 387 galaxies cataloged in UGC
with those given in PANBG. The result is shown in Fig. 3.
The average difference, standard deviation and maximum difference
of $b/a$ and $PA$ between the two sets were computed. The values are 
0.06(7.5\%), 0.07, and 0.54($67.5\% $) in $b/a$, and 
7.2$^{\circ}$(8.0\%), 15.6$^{\circ}$, 
 and 80.0$^{\circ}$(88.9$\%$) 
in $PA$.
They found that for both $b/a$ and $PA$ the difference increases
for more face-on galaxies. 
KO also found that most of 71 galaxies cataloged in the UGC with measured
diameters but without $PA$ are face-on galaxies. They concluded that
the selection effect against face-on galaxies in the UGC is too large 
for them to be  
included in analyses that make use of $b/a$ and $PA$.

Hu et al. (1993) and HWSL also discussed selection effects 
in the UGC (see Fig. 4) based on their analysis of 310 disk 
galaxies in the Virgo area, which
will be discussed in more detail in section 7.2.
They found that UGC galaxies with $PA$ preferentially have high
inclination angles $i$;  90\% of them have $i>40^{\circ}$,
and no galaxies with $PA$ are found for $i<20^{\circ}$.
They also found that the number of UGC galaxies with $b/a$
but without $PA$ is about 20\% of the total, of which about
80 $\% $ have $i<40^{\circ}$.
The remaining 20\% (i.e., 13 galaxies) with $i$ between 40$^{\circ}$
and 70$^{\circ}$ 
are found to be mostly faint Sm/Im galaxies and peculiars,
i.e., 4 Sm, 6 Im, 1 Imp and 1 Sp with $B_{T}$ between 14.0 and 
18.1 mag., except one bright S0p (Hu et al. 1993; HWSL).
Thus, they concluded that in addition to the selection 
effect against nearly face-on galaxies, 
a second morphology-dependent selection effect
against faint galaxies with irregular shapes
is present in the UGC.
This indicates that incompleteness increases as a galaxy sample goes
fainter, particularly for those galaxies with irregular shapes like Sm, Im 
and peculiars.

Most studies based on UGC galaxies either did not include face-on 
galaxies without $PA$ given in the catalog. 
Others simply assumed some certain values  for the $PA$ of those galaxies.
For example, KS and FG assumed $PA = 0^{\circ}$, and
Godlowski (1994) and HWSL assigned
by computer simulation random
values to supergalactic $PA$ or $PA$. 
KS claimed that assigning $PA=0$ leads to errors of $\leq30^{\circ}$
in the $PA$ of the($\sim$20\%)  face-on galaxies in their sample. 
When surface photometry is available, $PA$ and $b/a$ are determined
usually by a least squares fit of an ellipse to the isophote
at a certain surface brightness level, e.g., 25 mag arcsec$^{-2}$ 
as in PANBG. 
The internal accuracy of the PANBG data were estimated 
as, $\sigma(PA_{25})\sim4.5^{\circ}$ (only for galaxies with $b/a<0.8$),
$\sigma(\rm{log}\: a_{25})=0.03$, and
$\sigma(b_{25}/a_{25})=0.04$ (PANBG). 
By using the data measured by surface photometry, the face-on problem 
could be mitigated.
However, the isophotes of galaxies with irregular shapes are not well 
fitted by an ellipse.
KO found 31 galaxies ($\sim 5\%$ of their total sample) seems 
to have large errors, but they claimed that their results 
were not affected by this.

\vspace{3mm}\noindent

(b) Effects of Measurement Errors and Other Uncertainties

\vspace{2mm}
In this section we discuss the effects resulting from the errors in the data 
and some approximation treatments.
Firstly, there is the Holmberg effect: a systematic difference between diameters 
and axial ratios visually measured on photographic plates, and those measured 
using photometer tracings or surface photometry (Holmberg 1975).
One must take this effect into account when one  uses visually estimated
diameters and axis ratios. One way to do so is to define a system 
of standard diameters from photometric measurements, and to 
reduce other visual measurements to the system.
Fouqu\'e and Paturel (1985) presented an improved standard 
system of 250 galaxies based on the catalog of 237 standard 
photometric diameters 
(Fouqu\'e and Paturel, 1983).
They presented the formulae to convert the visual diameters 
of galaxies given in large catalogs (ESO, UGC and MCG) 
to this system. 

FG examined the effect of random errors of the order of $0.1'$ in both 
$a$ and $b$, and those of $5^\circ$  
in $PA$. They 
obtained
quite similar results when these errors are
introduced independently or simultaneously.
Flin (1989) also analyzed the influence of errors in $PA$ and 
$b/a$. He pointed out that errors of $\Delta(b/a)=0.1$
and $\Delta{PA}=13^\circ$, combined with the flatness of the LSC, 
did not give rise to spurious anisotropy.
Flin and Godlowski (1989) took into account random errors 
of $\sim0.1'$ in $a$ and $b$ and
$\sim5^\circ$ in $PA$ for all samples, and assumed a random distribution for 
$PA$ or inclination angle $i$ for face-on galaxies.
These authors found that the observed effect did not change the results 
qualitatively, although a 
quantitative difference did appear.
FG and Godlowski (1993) concluded that the supergalactic 
$PA$ of face-on galaxies are distributed essentially random.  Godlowski (1994) 
claimed that angles $\theta$ 
and $\phi$ of face-on galaxies depend only weakly on the 
supergalactic $PA$. Based on their previous researches,
Godlowski (1994) claimed that assigning 
a random distribution to  the supergalactic $PAs$ of face-on galaxies would
be a good approximation.

There are more factors which can affect the
measurement accuracy, even if the data are derived from surface photometry.
For example, poor seeing makes the elongated image of galaxies
rounder, i.e, makes $b/a$ larger. Other effects which lead to
non-circular images such as telescope tracking error and
PSF anisotropy also affect $b/a$ and $PA$.
However, all these effects usually do not exceed a few arcsecs and are 
negligible for bright galaxies in the 
LSC discussed here.
For example, the limiting diameters of galaxies of the samples
discussed here are of the order of 1 arcmin;
$1.'0$ (UGC), $0.'9$ (ESO/U), and $1.'7$ 
(FGCP discussed in section 7.2; Fouqu\'e et al., 1992).
A random sampling showed that the typical value of diameters of
the PANBG galaxies with velocity smaller than 3000 km s$^{-1}$ 
and brightness near 13 mag is also about $1.'0$ ($0.'8-1.'3$). 
However, we must be aware that these effects
can become important when we analyze smaller and/or fainter 
galaxies in the LSC or distant galaxies outside of the LSC.

\subsection {Inclination Angle}

Inclination angle $i$ can be derived from equation (3), which 
is valid for oblate spheroids.
It is generally believed that disk galaxies are fast rotating oblate
objects and that their rotation axes are normal to their disk planes.
It is thus reasonable to use equation (3) for disk galaxies.

On the other hand, intrinsic shape and rotation of elliptical
galaxies are more complex. 
The twist of $PA$ observed in many elliptical galaxies suggests
that they do not possess an axis of rotational symmetry, but are
triaxial objects (Mihalas and Binney, 1981; Binney and Tremaine, 1987,
 and references therein).
Some ellipticals may be prolate spheroids rather than oblate spheroids.
It is hard to define a meaningful plane and an axis of rotating
symmetry (i.e., a SV) for triaxial objects. 
Furthermore, most bright ellipticals are known to be slow rotators
(Illingworth, 1977, Davies et al., 1983; Binney and Tremaine, 1987; 
Binney and Merrifield, 1998 and references therein). 
The origin of rotation of ellipticals may be different from 
that of disk galaxies.
When ellipticals were included in the analysis in previous
studies, their inclination angles were estimated by equation (3)
in the same manner as disk galaxies.
Considering the complexities arising from intrinsic shape and 
slow rotation, we recommend not to use elliptical galaxies in
the analysis of SVs.

For disk galaxies, the inclination angle $i$ is derived from the observed 
$b/a$ and intrinsic axial ratio $q_0$ through equation (3).
The observed values of $b/a$ were sometimes converted to a standard, 
photometric axial ratio using the formula given by Fouqu\'e and 
Paturel (1985) (e.g. Flin and
Godlowski, 1989;  Godlowski, 1994).
Flin and Godlowski (1989) discussed the 
Holmberg effect using their 2227 UGC and ESO/U galaxies.
They considered various cases;  $q_0=0$, and $q_0$ $\not=0$ with and
without the Holmberg effect taken into account.
They found the following: Neglecting the Holmberg effect results in 
a greater anisotropy than that when
it is allowed for. The smallest anisotropy
is found when $q_0$ $\not=0$ and the Holmberg effect is
neglected. However, the coefficients that characterize the
deviation from isotropy do not change significantly among all the
cases. 
Using 1275 UGC galaxies, FG also showed that the differences 
with  or without accounting for the Holmberg effect
(for $ q_0 \not= 0$ ) were statistically negligible. 

\subsection {Selection Effects on Galaxy Position and Inclination Angle}

Using simulations including $2\times10^{5}$ virtual galaxies, Aryal and
Saurer (2000) computed the distribution of polar and azimuthal angles
of galaxy rotation axes expected for the isotropic distribution.
They showed that the selection effects concerning positions
and inclination angle $i$ of galaxies 
may play an important role when samples are incomplete 
(e.g. limited sky coverage) (see Fig.5 below). 
They also pointed out that up to now no authors have used homogeneous 
data covering both the northern and southern 
hemisphere for their studies on SVs of galaxies:
FG mainly used the UGC data; Goldlowski (1993, 1994) added the 
data from ESO/U; KO, Yuan et al. (1997),
and HYWSL used the PANBG galaxies with $\delta > - 25^{\circ}$.
The correct treatment of face-on galaxies is also a crucial point 
when studying galaxy alignment.

Aryal and Saurer (2000) showed that the expected isotropic distribution curve
is cosine for the polar angle $\theta$ and a straight line for 
the azimuthal angle $\phi$ 
{\it only when} there is no selection effect in (L, B), 
supergalactic position angle $PA$ and
inclination angle $i$.
Their main conclusions on selection effects are as follows:
1, The $\theta$ distribution deviates from the cosine curve 
when some selection effect is present in the axis ratios, i.e., 
inclination angles $i$.
The way of deviations is dependent on B.
2, The distribution of $\phi$ is very much affected by selection
effects. Even if there is no selection effects on B and $i$,
selection effects on L make the distribution deviate from the
expected straight line. Selection effects
on $i$ also change the $\phi$ distribution diffferently
for different ranges of L.

As an example, we show in  Fig. 5 the difference between the expected 
isotropic distribution 
calculated by Aryal and Saurer (2000) as solid lines and the usually 
adopted one under the assumption of no selection effect
with dashed lines,
for the $\theta$ distributions of the member 
galaxies of the Virgo cluster (HWSL) and for the $\phi$ distributions 
of the field galaxies (HYSWL).
These authors claimed that since the selection effects on the position and the inclination 
angle are present in all available databases, proper simulations 
should be made to interpret the results. 

\section{The Results from Disk Galaxies}

In this section we will concentrate on galaxy alignments of 
disk galaxies, i.e., spirals and lenticulars obtained from modern databases.

\subsection{New database: Photometric Atlas of Northern Bright Galaxies,
and KO's study}

A new database PANBG became available in 1990, which is based on
homogeneous surface photometry of galaxies
observed with the 105 cm Schmidt telescope at Kiso Observatory.
PANBG gives photometry parameters, isophotes, luminosity profiles,
position angles, and axial ratios at 25 mag arcsec$^{-2}$ in the $V$
band for 791 galaxies brighter than 13 mag 
selected from the RSA
catalog. PANBG includes 85\% of RSA galaxies at $\delta > - 25^{\circ}$.
It represents the main part of bright galaxies in the LSC
and therefore can be used as a good sample 
for the study of
galaxy orientations in the LSC. The accuracy of the PANBG data has been
discussed in section 6.1 (a).

Using this uniform photometry database of galaxies, KO inspected the
orientation of SVs of  spirals and lenticulars with
$ {\it cz} \leq $ 3000 km s$^{-1}$. They divided the sample into
13 subsamples by the
following criteria: 

(1) distance from the LSC plane $|SGZ|$;

(2) projected distance $r$ on the LSC plane from the center of
the LSC (inner or outer region separation at  $r$ = 10h$^{-1}$Mpc); 

(3) Virgo cluster membership (galaxies located within a circle of 6$^{\circ}$ 
radius centered on M87); 

(4) absolute magnitude $M_V$; 

(5) size $D$ (diameter of the major axis); and 

(6) morphology $T$. 

Table 1 shows the statistics for the total sample and the subsamples.
The 'area' distribution is the number of galaxy SVs which point to
36 sky regions  divided in terms of the supergalactic coordinates (L, B).
Figure 6 shows the $\theta$ and $\phi$ distributions of SVs of the
total sample.

They found the following.
Firstly, since all the P($>x_\nu^2$) 
values for the $\theta$, $\phi$, and area 
distributions of the total sample exceed the critical value 
0.050 (Table 1),
they concluded that the distribution of SVs of the 618 sample
galaxies taken as a whole is isotropic at 95$\%$ confidence level.

Secondly, as seen 
in
Fig. 7, the $\theta$  
distributions
of SVs of the
$|SGZ|$ subsamples show a tendency that galaxies near the
LSC plane tend to have SVs parallel to the LSC plane while those
off the LSC plane tend to lay their SVs perpendicular to the LSC plane. They 
referred to this tendency as the $|SGZ|$ effect.
The $|SGZ|$ effect is almost the same effect that Godlowski (1994)
found in terms of the supergalactic latitude.

Thirdly, the galaxies in and near the core of the Virgo cluster
show a remarkable anisotropy in the $\phi$ distribution, and 
the galaxies of the total sample show a 
similar
tendency as well. 
The projections on the LSC plane of their SVs tend to point to the center of the
Virgo cluster, which confirms the claim by MacGillivray and Dodd
(1985b).
KO found that all of the subsamples shows a 
dip (more than 1 $\sigma$) 
below 
those expected from isotropy in $\phi=0^\circ-30^\circ$ 
(over at least two bins), 
and that more than a 4 $\sigma$ deviation
appears 
in both 2a (inner region) and 3a (Virgo member) subsamples.
In the $\phi$ distribution of the total sample
shown in Fig. 6b, a remarkable dip (2 $\sigma$) over 4 bins exists at 
$\phi=0^\circ-40^\circ$ ($\sim$ the direction of SGX axis) 
at right angles with the direction
to the center of the Virgo cluster ($\sim$ the direction of SGY axis).
This means that the projections on the LSC plane of SVs of the galaxies
tend to point to the Virgo center. 
In addition to the 
usually 
adopted isotropic $\phi$ distribution (solid line) mentioned above,
the expected isotropic
distribution with the selection effects taken into account is also 
shown by the dashed line in Fig. 6b, which was kindly provided by 
Aryal and Saurer (2004a).
Compared with this expected isotropic curve, the dip near
$\phi=0^\circ $ is still clearly visible ($>1\sigma$).
The conclusion on SV projection for the KO total sample is valid
even if we consider the selection effects.

\subsection{Morphology Dependence of the Virgo Cluster Anisotropy}

Galaxy alignments in the Virgo cluster --the core of the LSC-- have been
well studied by many authors. However, the results derived are quite 
diverse, and some are even qualitatively inconsistent.
Some authors show that the distribution of SVs is random (e.g.,
Thompson 1976; Hoffman et al. 1989),
but others claim that it is anisotropic (e.g., MacGilivary et al.1982;
Helou and Salpeter 1982). Even in the anisotropic case, the
SVs show an excess in the parallel (e.g., KS) or in the perpendicular
(e.g., FG) direction to the LSC plane, or even bimodal (e.g., Adams et al. 1980; 
KO).
These inconsistencies may be due to the difference in sample size, 
sampling criteria, 
method and database adopted, and/or
accuracy of the data and selection effects, etc. KO analyzed
the disk galaxies of the Virgo cluster using the PANBG based
on homogeneous surface photometry. However, the total 
number of their sample galaxies was only 74.
Therefore, HWSL made a study of the Virgo Cluster with the new
method and a large sample. 

The basis of HWSL study was the catalog of 310 disk galaxies
in the Virgo cluster compiled by Hu et al. (1993) 
using the three catalogs, i.e., the catalog of 2096 galaxies in 
the Virgo cluster region (Binggeli et al., 1985; hereafter VCC), 
which is based on the large scale ($10''.8 {\rm mm}^{-1}$)
and wide field (2.3 deg$^{2}$ ) photographic survey of the 
Virgo area by the 2.5-meter du Pond telescope 
at Las Campanas, the catalog of groups and group members
within 
80 Mpc ($H_0$= 75 km s$^{-1}$ Mpc$^{-1}$)) compiled 
with the revised hierarchical algorithm 
(Fouqu\'e et al., 1992; hereafter FGCP), and UGC.
The HWSL sample is essentially complete down to a  limiting diameter
1.0 arcmin (POSS blue plate) or a limiting magnitude $m=14.5$ mag  
for VCC galaxies, and a blue isophotal diameter D$_{25}$  at 
25 mag arcsec $^{-2}$  larger than 100 arcsec or an estimated 
limiting magnitude $m=14.2$ mag for FGCP galaxies.  
This data set is the largest one so far, and the corresponding  
catalog is the largest one ever used in the study of
galaxy orientations in the Virgo cluster.

HWSL restricted their sample to disk galaxies and 
separated the sample galaxies into two broad
morphological types, spirals (S) and lenticulars (S0).
Morphological types are taken from VCC, which is based on 
the data of higher image resolution than UGC.
PAs are taken from UGC. 
The diameters and axis ratios, 
log $D_{25}$ and log $R_{25}$, are taken from VCC;
these values were originally given by de Vaucouleurs 
and Pence (1979), and Binggeli et al. (1984). 

The number of galaxies with/without
$PA$ given in UGC is 178/51 for spirals and 67/14 for lenticulars.
HWSL extracted from the sample three partially overlapping 
data sets, SS'E, VCC $6^{\circ}$ and VI. The SS'E set contains the member 
galaxies of S, S' and E clouds of the Virgo I cluster.
The VCC $6^{\circ}$ set contains all certain and possible VCC cluster 
members with $PA$ and/or diameters given in UGC which are located
within $6^{\circ}$ from the cluster center.
The VI set consists of the FGCP galaxies belonging to the Virgo I 
cluster, M group and W cloud.

The $\theta$ distributions of SVs of HWSL galaxies with $PA$ given
in UGC for S, S0, and S+S0 types are shown
in Fig.8a, 8b, and 8c, respectively.
It is remarkable that two humps and a dip are seen in all the 
three panels at nearly the same $\theta$. 
One of the two humps are at low ($\theta\sim30^{\circ}-50^{\circ}$) 
and the other is at high ($\theta\sim70^{\circ}-80^{\circ}$) ranges,
though the high-$\theta$ hump is hardly visible in S0.  
They correspond to excesses of galaxies, with
respect to the isotropic distribution, whose SVs tend to be nearly
parallel and nearly perpendicular to the LSC plane.
The dip is seen at $\theta\sim0^{\circ}$. 
Features of these humps and the dip are summarized in Table 2.
HWSL found that 
the total number of galaxies responsible for the dip was roughly
equal to the number of galaxies without $PA$ given in UGC for
all the cases.
When they assigned random values to $PAs$ of those galaxies without 
$PA$ but with diameters given in UGC, the two humps remained 
essentially unchanged while the dip almost disappeared leaving
a remnant narrow dip at $\theta<10^{\circ}$ as shown in Fig.9
for the data set VI.
This remnant dip can be understood as the result of the 
selection effect of the Virgo cluster from the results of  
the 
simulation (see Fig.6a for the spirals (VI); Aryal and Saurer, 2000).

In summary, HWSL found that the distribution of SV orientations 
of disk galaxies in the Virgo cluster shows anisotropy in the $\theta$
distribution, and that
there is also a discernible anisotropy in the $\phi$ distribution
in the sense that projection onto the LSC plane of SVs tends
to point to the center of the Virgo cluster. Earlier findings by 
MacGillvray and Dodd (1985b) and KO are confirmed.
HWSL  also showed that the anisotropy is dependent on morphology.
Spirals show two humps in both the high and low-$\theta$ ranges
while lenticulars show the low-$\theta$ hump only. This indicates
that lenticulars show little excess of SVs perpendicular to 
the LSC plane. In the $\phi$ distribution, the dip near
the direction perpendicular to the center of the Virgo cluster 
appears to be deeper for lenticulars
than for spirals.
This means that the effect of SVs pointing towards the center 
of the Virgo cluster may be more pronounced for lenticulars than 
for spirals.

Note that Wu et al. (1997, 1998) found a similar
morphology dependence in the Coma
cluster (a rich cluster of galaxies) as that found in the Virgo cluster
(a loose cluster of galaxies), suggesting that the morphology dependence
of the orientation of disk galaxies in clusters is independent of the
richness of clusters (Hu et al., 1996).

\subsection{'Field' Galaxies as the Probe of Formation Epoch}

Fuller, West, and Bridges (1999) showed that the brightest galaxies in
some poor clusters, like their counterparts in richer Abell clusters,
are preferentially aligned with the principal axes of their host clusters
as well as the surrounding distribution of nearby Abell clusters.
They suggested that these alignments, independent of cluster
richness, are most likely produced by formation of the brightest cluster
galaxies by anisotropic infall along filamentary structures.

However, galaxies in the high-density
regions of clusters or superclusters may not be the best objects to probe
the condition at the formation epoch because the post-formation dynamical
effects such as merging, tidal effects, and other gravitational effects
can disturb the original distribution of galaxy orientations.

The fact that the SVs of galaxies in the Virgo cluster
tends to point to the cluster center (e.g., MacGillivray and Dodd, 1985b;
KO, and HWSL) may be evidence for such dynamical effects.
Galaxies in low-density regions in the LSC, i.e. those which do not
belong to any cluster or group may preserve the fossil nature of the
SV orientations better. We will refer to these as 'field' galaxies.
It is interesting to see whether the morphology dependence of the 
orientation of disk galaxies found in Virgo cluster also exists in field 
galaxies.

The samples studied by KO included the members of groups and clusters.
Hu et al. (1998; hereafter HYSWL) constructed a new sample of
bright field disk galaxies in the LSC from the PANBG database by
rejecting the members of groups which consist of more than three members
(Fouqu\'e et al., 1992).
This sample comprised 220 field disk galaxies brighter than 13th
mag and radial velocities $V\leq$3000 km s$^{-1}$. HYSWL divided the
sample into four subsamples according to morphology;  S0, Sa-bc, Sc-m,
and Sa-m. Table 3 shows their results of $\chi^2$ test for $\theta$,
$\phi$ and 'area' distributions. The $\theta$
distributions of the subsamples are presented in Fig. 10.

HYSWL found the following. 
Firstly, the orientations of SVs of the
field disk galaxies are significantly different from those 
of 618 PANBG disk galaxies as a whole 
in both the $\theta$ and $\phi$ distributions.
The distribution of the SVs of field galaxies(total sample) shows a weak tendency 
parallel to the LSC plane. 
It is worthwhile to 
mention that, if we take into account the selection effect pointed by
Aryal and Saurer (2000), the  projections of SVs of the field galaxies (total 
sample) onto the LSC plane are
isotropic (see Fig. 5a).
This is completely different from the cases of the LSC 
as a whole (e.g., total sample of KO), and the Virgo cluster
(e.g., KO and HWSL) 

Secondly, the distribution of SV orientations
of the field disk galaxies is also morphology dependent.
Lenticulars, which shows the most concentrated spatial distribution
toward the LSC plane, i.e., with the smallest mean $|SGZ|$
value, show a significantly anisotropic distribution of orientations.
Their SVs tend to align parallel to the LSC plane. Spirals, 
on the other hand,
in general do not show any significant alignment of their SVs with the LSC 
plane.
Fig. 10 shows clearly that the distribution of SVs is anisotropic for
lenticulars while it is isotropic for spirals. This is confirmed by the
the results of $\chi ^2$ test of $\theta$ distribution given in Table 3.

Third, the $|SGZ|$ effect claimed by KO is confirmed for field galaxies.
It may be largely due to the morphology dependence.

Based on these results, HYSWL suggested that lenticulars and
spirals might have different formation processes, and that
field galaxies might therefore be a better probe to detect information
about the formation of the LSC in the early universe. 

The results of $\chi^2$ tests of the $\theta$
and $\phi$ distributions are somewhat dependent on the bin size.
Yuan et al. (2000) used the K-S test to re-examine the SV
distributions of all the samples analyzed by HYSWL.
They confirmed the results of HYWSL that  the $\theta$ distribution 
of the field spirals is isotropic
while that of the
field lenticulars is anisotropic. 
The bin size of $10^\circ$ adopted by HYSWL gave the same
results as K-S test in the statistical sense.
The K-S test also revealed that the field spirals (Sa-bc galaxies)
and lenticulars (S0) do not originate from the same parent
distribution at the 96.83 $\%$ confidence level.

Yuan et al. (1997) constructed another sample of 302
galaxies within the LSC by rejecting the members of groups with at
least five members.  For this sample, the $\theta$ distribution
of neither lenticulars nor spirals shows a tendency for the 
SVs to be parallel to the LSC plane. The $\chi^{ 2}$ test of
$\theta$ distribution confirms that spirals and lenticulars
show an
isotropic distribution, a result different from HYSWL.
This suggests that the $\theta$ distribution of field
disk galaxies is sensitive to the criteria for field galaxies.

\subsection{Discussion}

The spatial orientation of disk galaxies in superclusters may give us some
clues on the formation processes at  an early stage of their 
formation.
The distribution of SVs of the field disk galaxies is
significantly different from that of cluster member galaxies.
This suggests that cluster member galaxies suffered
dynamical effects after formation, 
and changed their spin orientations collectively.

Field disk galaxies show 
a weak morphology-dependent tendency to have  their 
SVs parallel to the LSC plane. 
Lenticulars show a strong tendency, while spirals do not show such 
tendency.
The isotropic feature in the $\theta$ distribution
of SVs of the spirals looks to support the classical
bottom-up (or CDM) model, i.e., gravitationally
hierarchical clustering formation scenario of the LSC.
On the other hand, the anisotropic distribution for lenticulars,
which are spatially more concentrated towards the LSC plane,
is apparently consistent with the prediction of the classical
'pancake' model.

The tendency that lenticulars lay their SVs
parallel to the LSC plane was also observed for samples
other than the HYSWL sample which consisted of PANBG field
lenticulars. It was found, as the $|SGZ|$ effect, for a sample of
289 galaxies near the LSC plane extracted from 618 PANBG
galaxies (KO).
As we already mentioned, there are also several independent
studies of galaxy alignment in the LSC not based
on PANBG which show a similar tendency,
e.g., 1275 UGC galaxies with radial velocities smaller
than 2600 km s$^{-1}$ (FG);
2227 UGC and ESO/U galaxies with radial velocities smaller
than 2600 kms$^{-1}$ and another independent sample from
the NBG catalog (Goldlowski, 1993, 1994).
We therefore conclude that the 
tendency of lenticulars
to have their SVs parallel to the LSC plane has a firm
observational basis.

In order to understand the morphological dependence of orientation
of the bright isolated disk galaxies in the LSC, 
a morphology-dependent
hybrid theory of galaxy formation
might have to be invoked. At this point, however, we should keep
in mind that the results of orientation of the PANBG field galaxies
as discussed in HYSWL is based on only 220 field disk galaxies in
the LSC. More observations are  necessary
to confirm the results of the PANBG field galaxies.

\section{Summary and Prospects}

Progress has been made during the past decades in studying the
distribution of orientations of galaxies in the LSC :

\begin{enumerate}
\item The method used was upgraded from a two-dimensional analysis,
i.e. the analysis of the distribution of position angles and/or the
distribution of axial ratios independently, to a three-dimensional
analysis, where we analyze the direction of the spin vector 
(the normal to the galaxy plane) by taking account 
of both the position angle and axial
ratio simultaneously. In the three-dimensional analysis, however,
there are four possible solutions regarding the orientation of the SV
of a galaxy with a given position angle and a given axial ratio. This
ambiguity is caused by the fact that we do not know which 
side of the galaxy is closer to us. The availability of HI 
velocity fields removes  this ambiguity. 
However, there is not yet a large sample of LSC galaxies with such data.

\item The advent of the Uppsala General Catalogue of Galaxies (UGC)
 and the ESO/Uppsala Survey of the ESO(B) Atlas (ESO/U) stimulated
the research in this field. Based on homogeneous surface photometry
of galaxies, the Photometric Atlas of Northern Bright Galaxies (PANBG) 
provided a database for further studies of the orientations
of bright galaxies in the LSC.

\item Selection effects on the location of galaxies 
on the sky and on the 
inclination angle often play an important role. Proper simulations
should be made to interpret the results.

\item When the LSC is seen as a whole, galaxy planes tend to align 
perpendicular to the LSC plane, i.e. their spin vectors are aligned parallel to 
the LSC plane.  The projections of spin vectors onto the LSC plane 
tend to point to the Virgo cluster center.
There is evidence indicating that these 
tendencies are dependent on morphology,
with lenticulars showing the most significant effect.
The $|SGZ|$ effect of galaxy orientation found for the PANBG 
galaxies of the LSC is mostly
due to this 
morphology
dependence and to the fact that lenticulars
are spatially more concentrated toward the LSC plane than spirals.

\item Projections of the spin vectors onto the LSC plane of Virgo cluster 
member galaxies, and those of the total LSC sample, tend to point to the 
Virgo cluster center.
The effect is more pronounced for lenticulars than for spirals.
This  suggests that member galaxies of 
the Virgo cluster
may have experienced the post-formation dynamical effects.

\item 'Field' galaxies, i.e., those which do not belong to groups
with more than three members, might be better objects to search for
information about the epoch of galaxy formation.
Orientations of spin vectors of the 
220 field disk galaxies are significantly
different from those of 618 PANBG disk galaxies as a whole. 
The $\theta$ distribution of SVs is 
anisotropic for field lenticulars 
while it is isotropic for  
field spirals. Field lenticulars exhibit a clear trend to 
have their spin vectors parallel to the LSC plane.

\end{enumerate}

It is important to increase the sample size of LSC galaxies,
especially 'field' LSC galaxies, available to the analysis.
HYSWL used only 220 PANBG field galaxies (selected
by $ {\it cz} \leq $ 3000 km  s $^{-1}$) brighter than 13 mag.
Going deeper by one magnitude with relevant velocity information
would result in a significant increase.
Accurate position angles and axial ratios derived by surface photometry
are essential to obtain reliable results.
The use of the non-parametric KS test would also improve the analysis.

Up to now no study has used homogeneous data covering both
northern and southern hemisphere for galaxies of the LSC.
It is attractive to extend PANBG, a database based on homogeneous
surface photometry of bright galaxies with $\delta > -25^{\circ}$,
to southern hemisphere to form a whole sky complete sample of bright 
galaxies in the LSC. Using existing databases to extend the study of the PANBG 
field galaxies to cover the whole sky may be an alternative way ahead in this 
regard.

The Westerbork Survey of HI in Spiral Galaxies (WHISP) is carrying
out observation of the distribution and velocity structure of neutral
hydrogen in several hundred bright spiral galaxies. These data will
be of great help in resolving the degeneracy of the orientations of
the spin vector and will provide us with a powerful database for a
research on the orientations of bright spirals in the LSC.

\vspace{1cm}
Acknowledgement

\vspace{0.5cm}
The authors wish to express their gratefulness to 
Prof. Lu, C. L., Wei, D. M., Feng, L. L.,
and Shu, C. G. for their helpful discussions. Special thanks go to  Prof.
Su, H. J., Liu, Y. Z., Lu, T., Zou, Z. L., van Albada, T. S., Paturel, G.,
Hamabe, M., Saurer, W., and Aryal, B. for their constructive suggestions and
kindly providing us with valuable data including PGC/LEDA database, SPIRAL
image processing 
software
and their most recent results.  FXH wishes to 
express his hearty
thanks to Prof. Nalikar, J. V., Kembhavi, A., and Cheng, K.S. for their
hospitalities during his visits in Inter-University Center for Astronomy and 
Astrophysics
(IUCAA), Pune, India, and in Dept. of Physics, The University of Hong Kong.
Finally, we are also greatly indebted to 
an anonymous referee
and Prof. Wamsteker, W. 
for their valuable comments
and recommendations on the early version of our manuscript,
which significantly improved the paper.

This work is supported by National Natural Science Foundation of China
(NNSFC; 19873018) and 
the Grant-in-Aid from the Ministry of
Education, Science, Sports, and Culture (11640228), Japan
and in part by
The Third World Academy of Sciences South-South
Fellowship/IUCAA, the University of Hong Kong, and NNSFC 10273007.

\clearpage

\centerline{\bf References}

\vsn
Adams, M.T., Strom, K.M. and Strom, S.E.: 1980, {\it Astrophy. J.}, {\bf 238}, 
445.

\vsn
Aryal, B., and Saurer, W.: 2000, {\it Astron. and Astrophys.},{\bf 364}, L 97.

\vsn
Aryal, B., and Saurer, W.: 2001a, {\it Astronomische Gesellschaft 
Abstract Series}, {\bf 18}, 218.

\vsn
Aryal, B., and Saurer, W.: 2001b, {\it Astronomische Gesellschaft 
Abstract Series}, {\bf 18},
63.

\vsn
Aryal, B., and Saurer, W.: 2004a, Private Communication.

\vsn
Aryal, B., and Saurer, W.: 2004b, {\it Astron. Astrophys}, 
{\bf 425}, 871.

\vsn
Aryal, B., and Saurer, W.: 2005a, {\it Mon. Not. R. Astron.Soc.}, 
{\bf 360}, 125.

\vsn
Aryal, B., and Saurer, W.: 2005b, {\it Astron. Astrophys}, 
{\bf 431}, 841.

\vsn
Aryal, B., and Saurer, W.: 2005c, {\it Astron. Astrophys}, 
{\bf 425}, 431.

\vsn
Bahcall, N. A.: 1988, {\it Ann. Rev. Astron. Astrophys}, {\bf 26}, 613.

\vsn
Binggeli, B., Sandage, A., and Tarenghi, M.: 1984, {\it Astron. J.}
{\bf 89}, 64.

\vsn
Binggeli, B., Sandage, A., and Tammann, G.A.: 1985, {\it Astron. J.}
{\bf 90}, 1681.(VCC)

\vsn
Binney, J., and Tremaine, S.: 1987, {\it Galactic Dynamics}, (Princeton:
Princeton University Press), Princeton, New Jersey. p.216, p.256. 

\vsn
Binney, J., and Merrifield, M.: 1998, {\it Galactic Astronomy}, (Princeton:
Princeton University Press),  New Jersey.  p.184, p.710.

\vsn
Bottineli, L.,  Gouguenheim, L., Paturel, G., and de Vaucouleurs, G. : 1983,
{\it Astron. Astrophys.},  {\bf 118},  4.

\vsn
Brown, F.G.: 1938, {\it Mon. Not. R. Astron.Soc.}, {\bf  98}, 218.

\vsn
Brown, F.G. :1939, {\it Mon. Not. R. Astron.Soc.}, {\bf 95}, 534.

\vsn
Brown, F.G.: 1964, {\it Mon. Not. R. Astron. Soc}, {\bf 127}, 517.

\vsn
Brown, F.G.: 1968, {\it Mon. Not. R. Astron. Soc.}, {\bf 138}, 527.

\vsn
Bukhari, F.A., and Cram, L.E.: 2003, {\it Astrophys Space Sci.}, 
{\bf 283}, 173.

\vsn
Cabanela, J.E., and Aldering, G.: 1998, {\it Astron. J.}, {\bf 116}, 1094.

\vsn
Cabanela, J. E., and Dickey, J.M.: 1999, {\it Astron J.}, {\bf 118}, 46.

\vsn
Danver, C.-G.: 1942, {\it Ann. Lunds Astron. Obs.}, No. 10.

\vsn
Davies, R.L., Efstathious,G., Fall,S.M., Illingworth,G., 
\& Schechter, P.L.: 1983. 
{\it Astrophys. J.}, {\bf 266}, 41.

\vsn
de Vaucouleurs, G.: 1953, {\it Astrophys. J.}, {\bf 58}, 30.

\vsn
de Vaucouleurs, G.: 1956, {\it Vistas in Astronomy}, {\bf 2}, 1584.

\vsn
de Vaucouleurs, G.: 1958, {\it Astrophys. J.}, {\bf 127}, 487.

\vsn
de Vaucouleurs, G.: 1960, {\it Soviet Astr. -AJ}, {\bf 3}, 897.

\vsn
de Vaucouleurs,G., and de Vaucouleurs, A.: 1964, {\it Reference 
Catalogue of Bright Galaxies}, (Austin: University of Texas Press).(RC)

\vsn
de Vaucouleurs, G., de Vaucouleurs, A., and Corwin, Jr. H. G.:
1976, {\it Second Reference Catalogue of Bright Galaxies},
(Austin: University of Texas Press).(RC2)

\vsn
de Vaucouleurs,  G.: 1978, {\it IAU Symp.}, 79, p.205.

\vsn
de Vaucouleurs, G., and Pence, W.D.: 1979, {\it Astrophys. J. Suppl.}, {\bf 
40}, 425.

\vsn
de Vaucouleurs, G.: 1981, {\it Bull. Astron. Soc. India}, {\bf 9},1.

\vsn
de Vaucouleurs, G.: 1982, {\it Astrophys. J.}, {\bf 253}, 520.

\vsn
Djorgovski, S.: 1987, {\it Nearly Normal Galaxies}, eds. S.M.Faber et al.
(New York, Springer), p.227.

\vsn
Flin, P.: 1989, {\it Morphological Cosmology}, (eds) Duerbeck, H. W., and Flin, 
P.,
Springer-Verlag, New York - Berlin.

\vsn
Flin, P.: 1995, {\it Comments on Astrophys.}, {\bf 18}, 81.

\vsn
Flin, P.: 1996, {\it Astron. and Astrophys. Trans. },{\bf 10},  p.153.

\vsn
Flin, P.: 2001, {\it Mon. Not. R. Astron. Soc.}, {\bf  325}, 49.

\vsn
Flin, P., and Godlowski, W.: 1986, {\it Mon. Not. R. Astron. Soc.},
{\bf 222}, {\bf 525}. (FG)

\vsn
Flin, P., and Godlowski, W.: 1989, {\it Sov. Astron.lett.},
{\bf 15(5)}, 374.

\vsn
Fouqu\'e, P. and Paturel, G.: 1983, {\it Astron. Astrophys. Suppl. Ser.}, {\bf  53},
 351.

\vsn
Fouqu\'e, P. and Paturel, G.: 1985, {\it Astron. Astrophys}, {\bf 150}, 192.

\vsn
Fouqu\'e, P., Gourgoulhon, E., Chamaraux, P., and Paturel, G.: 1992, 
{\it Astron. Astrophys}, {\bf 93}, 211.(FGCP)

\vsn
Fuller, T. M., West, M. J., and Bridge, T.J.: 1999, {\it Astrophys. J.}
{\bf 519}, 22.

\vsn
Garrido, J.L., Battaner, E., Sanchez-Saavedra, M.L., and Florido, E.: 1993, 
{\it Astron. Astrophys.}, {\bf 271}, 84.

\vsn
Godlowski, W.: 1993, {\it Mon. Not. R. Astron. Soc.}, {\bf 265}, 874.

\vsn
Godlowski, W.: 1994,  {\it Mon. Not. R. Astron.Soc.}, {\bf 271}, 19.
 
\vsn
Han, C., Gould, A., and Sackett, P.D.: 1995, {\it Astrophys. J.}, 
{\bf 445}, 46.

\vsn
Hawley, D.L. and Peebles, P.J.E.: 1975, {\it Astron . J.}, {\bf 80}, 477.

\vsn
Heidmann,J., Heidmann,N., and de Voucouleurs,G.: 1971,  {\it Memoirs of the 
Royal 
Astronomical Society}, {\bf 75}, 85. 

\vsn
Helou G., and Salpeter, E.E.: 1982, {\it Astrophys. J.}, {\bf 252}, 75.

\vsn
Herschel, W.: 1784, {\it Collected Works}, (London: R. S. and R. A. S., 1912),
{\bf 1}, 157.

\vsn
Herschel, W.: 1785, ibid., {\bf 1}, 223.

\vsn
Herschel, W.: 1802, ibid., {\bf 2}, 199.

\vsn
Herschel, W.: 1811, ibid., {\bf 2}, 459.,

\vsn
Hoffman, G. L., Lewis, B.M., Helou, G., Salpeter, E.E., and 
Williams, H.L.I.: 1989,{\it Astrophys. J. Suppl.}, {\bf 69}, 65.

\vsn
Holmberg, E.: 1937, {\it Lund Ann.}, {\bf 6}.

\vsn
Holmberg, E.: 1946, {\it Medd. Lunds Astron. Obs.}, Ser. VI, No.117.

\vsn
Holmberg, E.: 1958, {\it Medd. Lunds Astron. Obs.}, Ser. 2, No.136.

\vsn
Holmberg, E.: 1975, {\it Galaxies and the Universe}, Ed. Sandage,A., Sandage,M. 
and  Kristian, J. The University of Chicago Press. 1975. p.151.

\vsn
Hu, F. X., Wu, G. X., and Su, H.J.: 1993, {\it Publ. Purple Mountain 
Obs.}, {\bf 12}, 65.

\vsn
Hu, F.X., Wu, G.X., Su, H.J., and Liu Y.Z.: 1995, {\it Astron. Astrophys.},
{\bf 302}, 45. (HWSL) 

\vsn
Hu, F.X., Wu, G.X., Su, H.J., Liu Y.Z., and Yuan, Q.R.: 1996, 
{\it J. Korean Astron. Soc.}, {\bf 29}, S53.

\vsn
Hu, F.X., Yuan, Q.R., Su, H.J., Wu, G.X., and Liu Y.Z.: 1998, 
{\it Astrophys. J.}, {\bf 495}, 179. (HYSWL)

\vsn
Illingworth, G. 1977, {\it Astrophys. J. Letters}, {\bf 218}, L43.

\vsn
Jaansite, J., and Saar, E.: 1977, {\it Tartu Obs.}, preprint A-2.

\vsn
Jaaniste, J., and Saar, E.: 1978, {\it IAU Symp.}, {\bf 79}, 448.

\vsn
Kapranidis, S., and Sullivan,W.T., III.: 1983, {\it Astron. Astrophys.},
{\bf 118}, 33. (KS)

\vsn
Karachentsev, I.D., Karachentseva, V.E., Parnovsky, S.L., Kudrya Yu, N:
1993, {\it Astron. Nachr.}, {\bf 314}, 97.

\vsn
Kashikawa , N., and Okamura, S.: 1992, {\it Publ. Astron. Soc. Japan},
{\bf 44}, 493. (KO)

\vsn
Kodaira K., Okamura, S. Ichikawa, S., Hamabe, M., and Watanabe, M.: 1990,
{\it Photometric Atlas of Northern Bright Galaxies}
 (Tokyo: Univ. Tokyo Press). (PANBG)

\vsn
Kristian, J.: 1967, {\it Astrophys. J.}, {\bf 147}, 864.

\vsn
Lauberts, A: 1982, {\it The ESO/Uppsala Survey of the ESO(B) Atlas},
(Munchen: ESO). (ESO)

\vsn
 Lundmark, K.: 1927, {\it Studies of Anagalactic Nebulae}, Medd. 
Uppsala Obs.,  No.30.

\vsn
MacGillivray,H.T., Dodd, R.J., McNally, B.V., and Corwin, H.G.: 1982,  
{\it Mon. Not. R. Astron. Soc.}, {\bf 198}, 605.

\vsn
MacGillivray, H.T., and Dodd, R.J.: 1985a, {\it Astron. Astrophys.}, 
{\bf 145}, 269.

\vsn
MacGillivray, H.T., and Dodd, R.J.: 1985b, {\it Proc. ESO workshop on the
Virgo Cluster of Galaxies}, eds. Richter, O.G., and Binggeli, B.
(Garching, ESO), p.217.

\vsn
Mihalas, D., and Binney, J.: 1981, {\it Galactic Astronomy}, 
(W. H. Freeman and Company), San Francisco. p.329.

\vsn
Nilson, P.: 1973, {\it Uppsala General Catalogue of Galaxies},
Nova Acta Uppsala University. Ser. V:A, Vol.1.(UGC)

\vsn
Oort, J. H.: 1983, {\it Ann. Rev. Astron. Ap.}, {\bf 21}, 373.

\vsn
Parnovsky, S.L., Karachentsev, I.D., and Karachentseva, V.E.: 1994,
{\it Mon. Not. R. Astron. Soc.}, {\bf 268}, 665.

\vsn
Reaves, G.: 1958, {\it Publ. Astron. Soc. Pacific}, {\bf 70}, 461.

\vsn
Reinhardt, M., and Roberdts, M.S.: 1972, {\it Astrophys. Letters}, 
{\bf 12}, 201.

\vsn
Reinmuth, K.: 1926, {\it Die Herschel Nebel}, Veroff. Sternw. 
 Heidelberg Bd. 9.

\vsn
Reynolds, J.H.:  1920, {\it  Mon. Not. R. Astron.Soc.}, {\bf 71}, 129.

\vsn
Reynolds, J.H.: 1922, {\it  Mon. Not. R. Astron.Soc.}, {\bf  82}, 510. 

\vsn
\vsn
Sandage A., and Tammann , G. A.: 1976, {\it Astrophys.J.}, {\bf 207}, L1.

\vsn
Sandage A., and Tammann , G. A.: 1981, 1987, {\it A Revised Shapley-Ames 
catalog of bright galaxies} (Washington, D.C.: Carnegie Inst.). (RSA)

\vsn
Shapley, H, and Ames A.: 1932, {\it Harvard Obs. Ann.}, {\bf 88(4)},
43 (SA).

\vsn
Sugai, H., and Iye, M. 1995, {\it MNRAS}, {\bf 276}, 327.

\vsn
Thompson, L.A. : 1973, {\it Publ. Astron. Soc. Pacific}, {\bf 85}, 528.

\vsn
Thompson, L.A. : 1976, {\it Astrophys. J.}, {\bf 209}, 22.

\vsn
Tully, R.B.: 1982, {\it Astrophys. J.}, {\bf 257}, 389.

\vsn
Tully, R.B, and Fisher, J.R.: 1987, {\it  Nearby Galaxies Atlas},
(Cambridge: Cambridge University Press).

\vsn
Tully, R.B.: 1988, {\it Nearby Galaxies Catalog}
(Cambridge: Cambridge Univ. Press). (NBG)

\vsn
Wu, G.X., Hu, F.X., Su, H.J., and Liu Y.Z.: 1997, 
{\it Astron. Astrophys.}, {\bf 323}, 317.

\vsn
Wu, G. X., Hu, F.X., Su, H. J. and Liu, Y.Z.: 1998, 
{\it Chin. Astron. Astrophys.}, {\bf 22}, 17.

\vsn
Yahil, A., Sandage, A., and Tammann, G.A.: 1980, {\it Astrophys. J.}
{\bf 242}, 448.

\vsn
Yuan, Q.R., Hu, F.X., Su, H.J., and Huang, K.L.: 1997, 
{\it Astron. J.}, {\bf 114}, 1308.

\vsn
Yuan,Q. R., Wang, L. and Shen, C. X.: 2000, {\it Journal of Nanjing Normal 
University(Natural Science edition)}, {\bf 23}, No.2. 47. (in Chinese)

\clearpage
\noindent
Table 1. Statistics for the total sample and the subsamples (Kashikawa 
and Okamura, 1992).$^1$\\ 
\begin{center}
\begin{tabular}{rclllllll}
\hline
  & & &\multicolumn{2}{c}{$\theta$ (9bins)} &\multicolumn{2}{c}{$\phi$ 
(18bins)}&\multicolumn{2}{c}{area(36bins)}\\
 &              &N   &$x_\nu^2$    &P($>x_\nu^2$) &$x_\nu^2$    &P($>x_\nu^2$) 
&$x_\nu^2$    &P($>x_\nu^2$)\\
\hline
         &Total sample  &618 &1.910 &0.054  &1.441 &0.106  &1.113 &0.297\\
\hline
         &Subsamples            & & & & & & & \\
(1)-1a   &$|SGZ|<2h^{-1}$ Mpc     &289 &2.543 &$0.009^*$ &0.787 &0.710 &0.909 
&0.621\\
   -1b   &$|SGZ|\geq$$2h^{-1}$ Mpc &329 &2.114 &$0.031^*$ &1.015 &0.438 &1.211 
&0.183\\
(2)-2b  &Inner region           &292 &0.871 &0.541 &1.687 &$0.038^*$  &1.831  
&$0.002^*$\\
   -2b   &Outer region           &326 &2.014 &$0.041^*$ &0.632 &0.870 &0.809 
&0.779\\
(3)-3a   &Virgo member           &74  &2.444 &$0.012^*$ &2.784 &$0.000^*$ &1.657 
&$0.009^*$\\
   -3b   &Non-member             &544 &2.379 &$0.015^*$ &0.925 &0.544 &0.947 
&0.556\\
(4)-4a   &Mag.~~~~~$Mv<-$$19+5$log$h$    &338 &2.730 &$0.005^*$ &1.307 &0.176 
&1.029 &0.422\\
   -4b          &~~~~~~~~~~~$Mv\geq-$$19+5$log$h$ &280 &0.779 &0.622 &0.824 
&0.667 &1.123 &0.284\\
(5)-5a  &\multicolumn{1}{l}{Size$~~~~~~D<7h^{-1}$ Kpc} &276 &1.040 &0.403 &1.101 
&0.345 &1.087 &0.333\\
   -5b          & ~~~~~~$D\geq7h^{-1}$ Kpc&342 &3.648 &$0.000^*$ &1.276 &0.197 
&1.176 &0.220\\
(6)-6a   &\multicolumn{1}{l}{Morph. $T=-3--1$}&122 &1.186 &0.303 &0.887 &0.590 
&1.627 &$0.011^*$\\
   -6b          &$T=0-4 $           &271 &3.659 &$0.000^*$ &1.000 &0.454 &1.024 
&0.430\\
   -6c          &~$T=5-10$           &225 &0.311 &0.962 &0.800 &0.695 &1.211 
&0.183\\
\hline
\end{tabular}
\end{center}
1) N is the number of sample. Asterisks mark the P($>x_\nu^2$) value 
which is smaller 
than the critical value (0.05), indicating the anisotropic distribution.

\vspace{1truecm}
\noindent
Table 2. The $\theta$ distribution of SV of galaxies of the Virgo cluster.
(Hu et al., 1995)
\vspace{1truecm}

\begin{center}
\begin{tabular}{lllll}
\hline
morphological& &(1)dip &(2)hump &(3)hump\\
type & &at $\theta\sim0^{\circ}$ & at low $\theta$ & at high $\theta$\\
\hline
spirals (S) & $\theta$   range & $0^{\circ}-30^{\circ}$ &
 $30^\circ-50^\circ$ & $70^\circ-80^\circ$\\
     &Deviation& $\sim3\sigma$ & $\sim1-2\sigma$ & $\sim1-2\sigma$\\
lenticulars (S0)   & $\theta$   range & $0^\circ-10^\circ$ & 
$20^\circ-50^\circ$ & \\
     &Deviation& $>4\sigma$ & $\sim1-2\sigma$ & \\
S+S0 & $\theta$   range & $0^\circ-20^\circ$ & $30^\circ-50^\circ$ &
$70^\circ-80^\circ$\\
     &Deviation& $>4\sigma$ & $\sim2-3\sigma$ & $\sim1-2\sigma$ \\
\hline
\end{tabular}
\end{center}

\vspace{1truecm}
\noindent
Table 3. The results of $\chi^2$ test of $\theta$, $\phi$, and area 
distribution (Hu et al., 1998). 
\begin{center}
\begin{tabular}{lllrlrlll}
\hline
Data sets and & & &\multicolumn{2}{c}{$\theta$ (9 bins)} &
\multicolumn{2}{c}{$\phi$ (18 bins)} &
\multicolumn{2}{c}{Area (36 bins)}\\
T selection &   N &$<|SGZ|>$ &$\chi^2_\nu$ &P($>\chi^2_\nu$) &$\chi^2_\nu$
 &P($>\chi^2_\nu$) &$\chi^2_\nu$ &P($>\chi^2_\nu$)\\
\hline
S0(-3$\leq{T}\leq$0) &41 &4.25 $h^{-1}Mpc$ &2.231 &$0.023^b$ &1.736 &$0.030^b$ 
&1.514 &$0.026^b$\\
Sa-bc(1$\leq{T}\leq$4) &84 &7.31 &0.768 &0.631 &1.992 &$0.009^b$ &1.537 
&$0.022^b$\\    
Sc-m(5$\leq{T}\leq$9) &90 &5.56 &0.900 &0.510 &2.106 &$0.005^b$ &1.519 
&$0.025^b$\\
Sa-m(1$\leq{T}\leq$9) &174 &6.40 &1.341 &0.218 &3.426 &$0.000^b$ &2.218 
&$0.000^b$\\
Total(-3$\leq{T}\leq$10)&220 &5.91 &1.255 &0.262 &4.520 &$0.000^b$ &2.932 
&$0.000^b$\\
\hline
\end{tabular}
\end{center}
$^b$ : P($>\chi_\nu^2$) $<$ 0.05 is regarded as the anisotropic distribution.

\clearpage

{\bf Figure Captions}
\noindent

\vspace{1cm}
Fig.1. Schematic illustration of angles $\theta$ and $\phi$ which defines
the orientation of the spin vector of a galaxy in the LSC. 
L and B are the supergalactic coordinates. 
SGX is directed towards the point L=0$^{\circ}$, B=0$^{\circ}$, and the 
SGZ towards the north super-galactic pole, B=+90$^{\circ}$. 
SGX and SGY are in the LSC plane. Angle $\theta$ is the polar angle between 
the galaxy spin vector and the LSC plane. Angle $\phi$ is the azimuthal 
angle between the projection on the LSC plane of the galaxy spin vector 
and the SGX-axis. 
The hatched ellipse represents a projection of the galaxy shape onto 
the sky, which we observe. 
Angle $P$ is the position 
angle in the supergalactic coordinate. (Kashikawa and Okamura, 1992) 

\vspace{1cm}
Fig.2. The observed (solid line) and the theoretical (dashed line) 
isotropic distributions of the angle $\delta$
(a), and angle $\eta$ (b) for the whole sample of galaxies.
(Godlowski, 1994)

\vspace{1cm}
Fig.3. Comparison of the axial ratio and the position angle
between UGC and PANBG for 387 galaxies.
(a) $b/a$(UGC) versus $b/a$(PANBG), and
(b) $PA$(PANBG)$-PA$(UGC) versus $b/a$(PANBG).
(Kashikawa and Okamura, 1992) 

\vspace{1cm}
Fig.4. The number of galaxies versus the inclination angle $i$ 
of 310 disk galaxies in the Virgo area with $PA$ and/or diameters
given in UGC. 
The inclination angle $i$ was estimated from log R$_{25}$,
the major to minor axial ratio at 25 $B$ mag arcsec $^{-2}$ 
taken from Binggeli et al. (1985), assuming the intrinsic 
axial ratio $q_0$ = 0.2.
The number of galaxies with UGC $PA$ (dash line), those
with UGC diameters but without UGC $PA$ (dotted line), 
and the total (solid line) are shown.
(Hu et al., 1993).

\vspace{1cm}
Fig.5 (a) The $\theta$ distribution of the spirals (the VI set) of 
the Virgo cluster.  A random distribution of $PAs$ is assumed for
the galaxies without $PA$ given in UGC (HWSL); 
(b) The $\phi$ distribution of the field galaxies (the total sample;
HYSWL). In both panels, the solid lines
are the expected isotropic distributions resulted from the simulations
by Aryal and Saurer (2000) while the dashed lines are isotropic
distributions without any selection effects taken into account.

\vspace{1cm}
Fig.6. The $\theta$ distribution (a), and $\phi$ distribution (b)
of the total sample of 
(Kashikawa and Okamura, 1992).
Statistical 1$\sigma$ error bars 
and the isotropic distribution without any selection 
effect is shown by the solid line. 
$\phi=0^\circ$ corresponds to +SGX
direction.
The expected isotropic distribution with the selection
effects taken into account is shown by the dashed line 
in panel (b), which was kindly 
provided by Aryal and Saurer (2004a). 

\vspace{1cm}
Fig.7  The $\theta$ distributions of the subsamples 
of (Kashikawa and Okamura, 1992): 
(1a) $|SGZ| < 2h^{-1}$Mpc, (1b) $|SGZ| \geq 2h^{-1}$Mpc.
Statistical 1$\sigma$ error bars are shown. The solid lines 
are the isotropic distributions without any selection effect.

\vspace{1cm}
Fig.8.  The $\theta$ distributions of SVs of lenticulars (a), 
spirals (b) and lenticulars plus spirals (c) with 
$PA$ given in UGC for various data sets of the Virgo cluster. 
The isotropic distribution without any selection effect 
is shown by the solid line. 
Statistical 1$\sigma$ error bars are indicated. (Hu et al.,1995)

\vspace{1cm}
Fig.9. The $\theta$ distributions of SVs of 
lenticulars (a), spirals (b) and lenticulars plus spirals (c) 
(Set VI) of the Virgo cluster when random values are assigned to $PAs$ 
of those galaxies without $PA$ given in UGC. (Hu et al., 1995).

\vspace{1cm}
Fig.10.  The $\theta$ distributions of SVs of the 'field' galaxies 
for S0(a), Sa-bc(b), Sc-m(c), 
and Sa-m(d) subsamples. 
Statistical 1$\sigma$ error bars are shown. The solid lines 
are the isotropic distributions without any selection effect.
(Hu et al., 1998).
\end{document}